\documentclass[aps,pre,twocolumn]{revtex4}
\usepackage{ae}
\usepackage{graphicx}

\begin{document}  
\narrowtext

\noindent
{\bf Rejoinder on: Thermostatistics of Overdamped Motion of Interacting Particles}

In their Reply~\cite{An11} to our Comment~\cite{LePa11}  Andrade {\it et al.} state that 
we have ``chosen to categorically dismiss 
their elaborate 
and solid conceptual approach without employing any concepts or tools from Statistical Mechanics''.  In this
Rejoinder we will show that contrary to the statement above, approach employed by  Andrade {\it et al.}
is neither ``solid'' nor ``sound''.  Unfortunately, because of the one page restriction imposed by  
PRL we could not address all 
of the flaws of the original paper.  Therefore, we are grateful to Andrade {\it et al.}, for giving us an opportunity
to further elaborate on our Comment.  Indeed, in our Comment we have shown that at $T=0$ the model
studied by  Andrade {\it et al.}~\cite{An10} can be solved exactly.  In their Reply Andrade {\it et al.} argued that
the thermodynamic limit employed by us to obtain the exact solution was ``less physical'' than their simulation
in which $800$ vertices were confined in a strip with  periodic boundary condition in one direction
and an external parabolic potential in the other.  To what extent this 
is more realistic than our thermodynamic limit calculation is very questionable, but lets leave it at that.

\begin{enumerate}

\item  Let us first briefly address the point of thermodynamic limit. Andrade {\it et al.} surely realize that
statistical mechanics, strictly speaking, is valid only in the thermodynamic limit.  Without this limit
there is no equivalence between different ensembles.  For particles confined inside a parabolic potential well
there are two ways that one can perform the thermodynamic limit (a) one can scale the vertex charge $q$ so that
it is proportional $1/\sqrt{N}$ or alternatively (b) one can leave q=1 and scale the confining potential so that its
stiffness $\alpha \propto N$.  In  either case the reduced one-particle distribution function will then converge
to the same function in  the limit,  $N \rightarrow \infty$.  The way that Andrade et al. performed their simulations,
there is no thermodynamic limit and for each different N the distribution will be different.  In thermodynamic
limit, the model of  Andrade {\it et al.} can be solved exactly for any $T$.  In the
Comment, we only presented the solution for $T=0$, but it can easily be extended to any $T$, showing that
the system of  Andrade {\it et al.} is always described by the usual Boltzmann-Gibbs statistical mechanics.

\item But lets us for a moment 
forget about the thermodynamic limit. In this case, mean-field is no longer exact and the particle
correlations must be taken into account.   
We first note that a finite number of particles
 confined in a periodic potential at $T=0$ will crystallize  -- of course, because
of the confinement, the crystal will have many defects.  It is, therefore,  meaningless to assign
a uniform functional form to density, which in reality will be made of a sum of delta-functions.  
It is like saying that NaCl
crystal is described by a constant density, with no Bragg peaks or anything related to the
crystal structure. At $T=0$ the only way to make sense of a differential equation 
like eq.(1) of Reply is if $N \rightarrow  \infty$.    

\item But even allowing the incorrect coarse-graining procedure of Andrade {\it et al.}, 
we still see that the density distribution will not
be given by a q-exponential. A  q-exponential is a smooth function, so that the density
must  go continuously to zero.  This is not the case for the system of  Andrade {\it et al.}.
In our Comment we saw that the particle density drops discontinuously to zero.  
For finite $N$, the particle correlations will change
slightly the form of the distribution function from the one obtained in the thermodynamic limit, but they will not
change the fact that there is a discontinuity in  density.   
This can be clearly seen from the simulation of Andrade {\it et al.} presented
in Fig.1 of their Reply.  The figure clearly 
shows a discontinuous jump in density.  Because of the way that Andrade {\it et al.} performed the binning 
to construct their histogram,  the jump has been smoothed out somewhat.   
We have rerun the simulations of  Andrade {\it et al.} for {\it exactly} the same parameters and
have plotted the data in Fig. 1.  The figure clearly shows a discontinuity 
in density, after the leftmost and the rightmost points of the histogram the density
is identically zero.  This discontinuity is predicted by our theory, and is
absent within the Tsallis formalism.    
We conclude again that the particle distribution at $T=0$ has nothing to do
with Tsallis statistics.
\begin{figure}[h]
\begin{center}
\includegraphics[width=7cm]{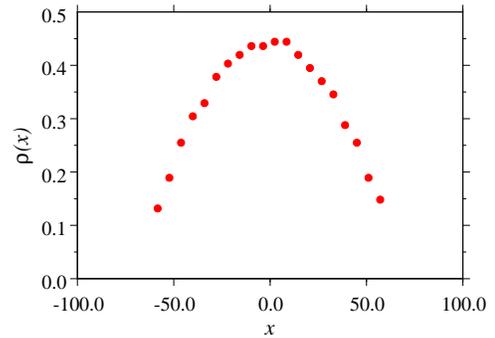}
\end{center}
\caption{The density distribution at $T=0$ for $N=800$ inside the parabolic potential well with $\alpha=10^{-3} f_0$.
The simulation is identical to the one done by  Andrade {\it et al.}.  After the two extreme points on the left and 
on the right, the density is identically zero.}
\end{figure}

\item  But again let us give a benefit of the doubt to Andrade {\it et al.} and ask the authors 
the following question.  For $T=0$ it is very simple to find the position of mechanical equilibrium for
N=1 or N=2,3,4,5...10 etc.  We  can also  calculate 
the particle distribution in the thermodynamic limit, as was done in our Comment.  In all these cases, 
the equilibrium is simply determined 
by the force balance.  No statistics is involved.  Our  question is this: for
exactly what value of N do authors expect that Tsallis entropy will start determining 
the particle distribution at T=0? 

\item  Now suppose that we raise the temperature.  At some point the crystal will melt.  However, the fluid
state will still have strong correlations between individual particles.  There is more than a hundred years of 
history, going all the way back to the pioneering works of Debye, Ornstein, Zernike, and Kirkwood on how the particle 
correlations can be calculated.  There are thousands of liquid-state  theorists 
who have dedicated their lives to understanding statistical mechanics of liquids  --- all this work simply
ignored by  Andrade {\it et al.}.    
In fact the correlations and their effects on the density distribution can be calculated 
in the framework of the usual Boltzmann-Gibbs statistical mechanics. 
There are theories, such as the Hypernetted Chain Equation, Roger-Young Equation, SCOZA, the  
Density Functional Theory, etc.  which were derived to do exactly this. There is
no reason to introduce any fitting parameters through Tsallis entropy.  This does not lead to any 
new physical understanding, only to curve fitting.  In fact, in their paper {\it  Andrade {\it et al.} 
did not provide any reason or indication why they believe that the standard Boltzmann-Gibbs statistical mechanics will
not apply to the system of vertices studied by them}.  

\item In their Reply, Andrade {\it et al.} say that we did not question their non-linear Fokker-Planck 
equation (eq. 1 of Reply)
on which are based all of their arguments.  Indeed, because of the one page restriction we could not
get into  detailed discussion of the original paper, concentrating only on our exact solution for $T=0$.  
In fact, {\bf we do question} the use of this non-linear Fokker-Planck equation 
to study the dynamics of interacting particles. This is an incorrect equation to use for this system. 
It is well known that the dynamics of interacting particles is governed by the self-consistent 
Nernst-Planck equation.  
Again, Andrade {\it et al.} got lucky because
for the particles interacting through a modified Bessel function, the stationary solution (after adjusted
with the fitting parameter as was done by  Andrade {\it et al.}) is quite close to the correct solution of 
Nernst-Planck Equation.  Thus, the asymptotic dynamics derived using the incorrect Fokker-Planck equation
actually {\it looks} quite reasonable.
In fact it is simple to see that eq. (1)
used by  Andrade {\it et al.} is incorrect.  Using it  Andrade {\it et al.} derived the unlucky 
eq. 13 of their PRL, from which all the incorrect discussion followed.  From this equation  Andrade {\it et al.} 
concluded that the stationary
state for particles interacting by {\it any short-range force} is always parabolic  -- a Tsallis q-Gaussian 
of entropic index $\nu=2$.  From our exact calculation,
it is clear that the parabolic form is very special for the interaction potential of the modified Bessel form.
Other potentials will not lead to the parabolic form. 
We should recall that Andrade {\it et al.} claim that their theory applies to particles
interacting by an arbitrary short-range potential.   
As a simple demonstration that this is not true, we have simulated
a 1d systems of particles interacting through a short range 
potential $V(x)=\frac{q^2}{2} e^{-x^4}$.  In Fig.2, we see that once the system has relaxed to
equilibrium, the distribution is not the inverse parabola, but a much more complicated function. 
One should also note
the characteristic discontinuity in density, contradictory to Tsallis statistics.   We conclude again that 
the non-linear Fokker-Planck equation of Andrade {\it et al.} can neither  
account for the stationary state nor for the dynamics of approach to equilibrium.  
\begin{figure}[h]
\begin{center}
\includegraphics[width=7cm]{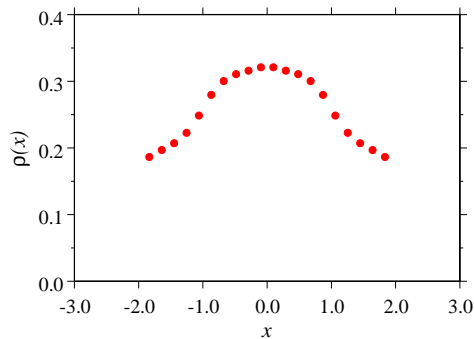}
\end{center}
\caption{The density distribution at $T=0$ for 
particles interacting by a short range potential  $V(x)=\frac{q^2}{2} e^{-x^4}$ confined in a parabolic trap.  
To get  better statistics
and faster run time, we studied a 1d model, with $N=1000$, $\alpha=0.05$. 
Note, however, that because of the periodic boundary condition the model of
Andrade {\it et al.} is also effectively 1d. Also note the discontinuous drop in density.}
\end{figure}

\item  Andrade {\it et al.} have misinterpreted the interparticle correlations present
in a finite system for existence of a new kind of entropy.  The fact that the system studied by
Andrade {\it et al.} is described by the usual Boltzmann-Gibbs statistical mechanics,
follows simply from the
study of velocity distribution in equilibrium.  In Fig.3 we
show a result of a microcanonical (constant energy) simulation.  The system starts from an arbitrary initial condition
and evolves in accordance with Newton's equations of motion.  After some time 
it relaxes to thermodynamic equilibrium. In Fig. 2 we plot the histogram of particle velocities together
with the 2d Maxwell-Boltzmann distribution, 
$$n(v)=\frac{2 v}{\langle v^2 \rangle} e^{-\frac{v^2}{{\langle v^2 \rangle}}}. $$   
We see a perfect agreement between the Boltzmann-Gibbs statistical mechanics and the simulations,
{\it without any adjustable parameters}.
In this example, the equilibrium temperature, $T= \langle v^2 \rangle/2 $, is very
low $T=0.3$ and  Andrade {\it et al.} claim that the system should be 
described by the Tsallis statistics.  
In fact we do not see any trace of 
q-exponentials at this or at any other temperature.  

\item In their Reply Andrade {\it et al.} state that ``neither the Maxwell-Boltzmann nor the Tsallis thermostatics
are contrary to classical mechanics theory''. We respectfully disagree with this statement.  Indeed
the Boltzmann-Gibbs statistical mechanics is in
full agreement with the classical mechanics.   However, there is not a single classical 
system of particles interacting by a short-range potential (for long-range forces see~\cite{LePa08} and the references therein)
that evolves to ``Tsallis'' equilibrium.  The model studied by  
Andrade {\it et al.} is precisely the case in point. 
Therefore, we stand by our original statement: 
``the density distribution 
of particles in contact with a reservoir at $T=0$ has nothing to do with the Tsallis statistics, and everything
to do with the Newton's Second Law.''

\end{enumerate}

\begin{figure}[h]
\begin{center}
\includegraphics[width=7cm]{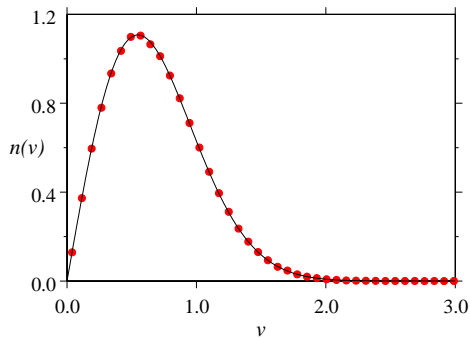}
\end{center}
\caption{Velocity distribution. Points are from molecular-dynamics simulation, for exactly the system
studied by Andrade {\it et al.}, $N=800$, $L_y=20$, $\alpha=10^{-3} f_0$.  Solid curve is the 
Maxwell-Boltzmann velocity distribution.  There are no adjustable parameters. The temperature is $T=0.3$.
At this temperature,  Andrade {\it et al.} claim that the system is described by the Tsallis statistics.  As is clear
from the graph, we see no trace of Tsallis q-exponentials in the velocity distribution function.}
\end{figure}

This work was partially supported by the CNPq, INCT-FCx, and by the US-AFOSR under the grant FA9550-09-1-0283.

\noindent
Yan Levin and Renato Pakter\\
Instituto de Física, UFRGS\\ CP 15051, 91501-970, Porto Alegre, RS,
Brazil


\begin{thebibliography}{4}
\bibitem{An11} J. S. Andrade, Jr., et al,  arXiv:1104.5036
\bibitem{LePa11} Y. Levin and R. Pakter, arXiv:1104.0697
\bibitem{An10} J. S. Andrade, Jr., et al., Phys. Rev. Lett. {\bf 105} 260601 (2010).
\bibitem{LePa08} Y. Levin, R. Pakter and T. N. Telles,  Phys. Rev. Lett. {\bf 100}, 040604 (2008);
T.N. Teles, Y.Levin, R. Pakter, and   F.B. Rizzato,  J. Stat. Mech. P05007 (2010);  R. Pakter and Y. Levin, arXiv:1012.0035 (2010)


\end{thebibliography}
\end{document}